\documentclass[sigconf]{acmart}

\usepackage{multirow}
\AtBeginDocument{
  }

\copyrightyear{2025}
\acmYear{2025}
\setcopyright{acmlicensed}\acmConference[SIGIR '25]{Proceedings of the 48th International ACM SIGIR Conference on Research and Development in Information Retrieval}{July 13--18, 2025}{Padua, Italy}
\acmBooktitle{Proceedings of the 48th International ACM SIGIR Conference on Research and Development in Information Retrieval (SIGIR '25), July 13--18, 2025, Padua, Italy}
\acmDOI{10.1145/3726302.3730211}
\acmISBN{979-8-4007-1592-1/2025/07}

\begin{document}

\title{Improving LLM-powered Recommendations with Personalized Information}

\author{Jiahao Liu}
\authornote{Equal contribution.}
\orcid{0000-0002-5654-5902}
\affiliation{
  \institution{Fudan University}
  \city{Shanghai}
  \country{China}}
\email{jiahaoliu21@m.fudan.edu.cn}

\author{Xueshuo Yan}
\orcid{0009-0005-2576-2292}
\authornotemark[1]
\affiliation{
  \institution{Fudan University}
  \city{Shanghai}
  \country{China}}
\email{23210240353@m.fudan.edu.cn}

\author{Dongsheng Li}
\orcid{0000-0003-3103-8442}
\affiliation{
  \institution{Microsoft Research Asia}
  \city{Shanghai}
  \country{China}}
\email{dongshengli@fudan.edu.cn}

\author{Guangping Zhang}
\orcid{0009-0001-9853-8268}
\affiliation{
  \institution{Fudan University}
  \city{Shanghai}
  \country{China}}
\email{gpzhang20@fudan.edu.cn}

\author{Hansu Gu}
\orcid{0000-0002-1426-3210}
\affiliation{
  \institution{Independent}
  \city{Seattle}
  \country{United States}}
\email{hansug@acm.org}

\author{Peng Zhang}
\orcid{0000-0002-9109-4625}
\authornote{Corresponding author.}
\affiliation{
  \institution{Fudan University}
  \city{Shanghai}
  \country{China}}
\email{zhangpeng\_@fudan.edu.cn}

\author{Tun Lu}
\orcid{0000-0002-6633-4826}
\authornotemark[2]
\affiliation{
  \institution{Fudan University}
  \city{Shanghai}
  \country{China}}
\email{lutun@fudan.edu.cn}

\author{Li Shang}
\orcid{0000-0003-3944-7531}
\affiliation{
  \institution{Fudan University}
  \city{Shanghai}
  \country{China}}
\email{lishang@fudan.edu.cn}

\author{Ning Gu}
\orcid{0000-0002-2915-974X}
\affiliation{
  \institution{Fudan University}
  \city{Shanghai}
  \country{China}}
\email{ninggu@fudan.edu.cn}

\renewcommand{\shortauthors}{Jiahao Liu et al.}

\begin{abstract}
Due to the lack of explicit reasoning modeling, existing LLM-powered recommendations fail to leverage LLMs' reasoning capabilities effectively. In this paper, we propose a pipeline called CoT-Rec, which integrates two key Chain-of-Thought (CoT) processes---user preference analysis and item perception analysis---into LLM-powered recommendations, thereby enhancing the utilization of LLMs' reasoning abilities. CoT-Rec consists of two stages: (1) personalized information extraction, where user preferences and item perception are extracted, and (2) personalized information utilization, where this information is incorporated into the LLM-powered recommendation process. Experimental results demonstrate that CoT-Rec shows potential for improving LLM-powered recommendations. The implementation is publicly available at \url{https://github.com/jhliu0807/CoT-Rec}.
\end{abstract}



\maketitle

\section{Introduction}
\begin{figure}[t]
  \centering
  \includegraphics[width=0.99\linewidth]{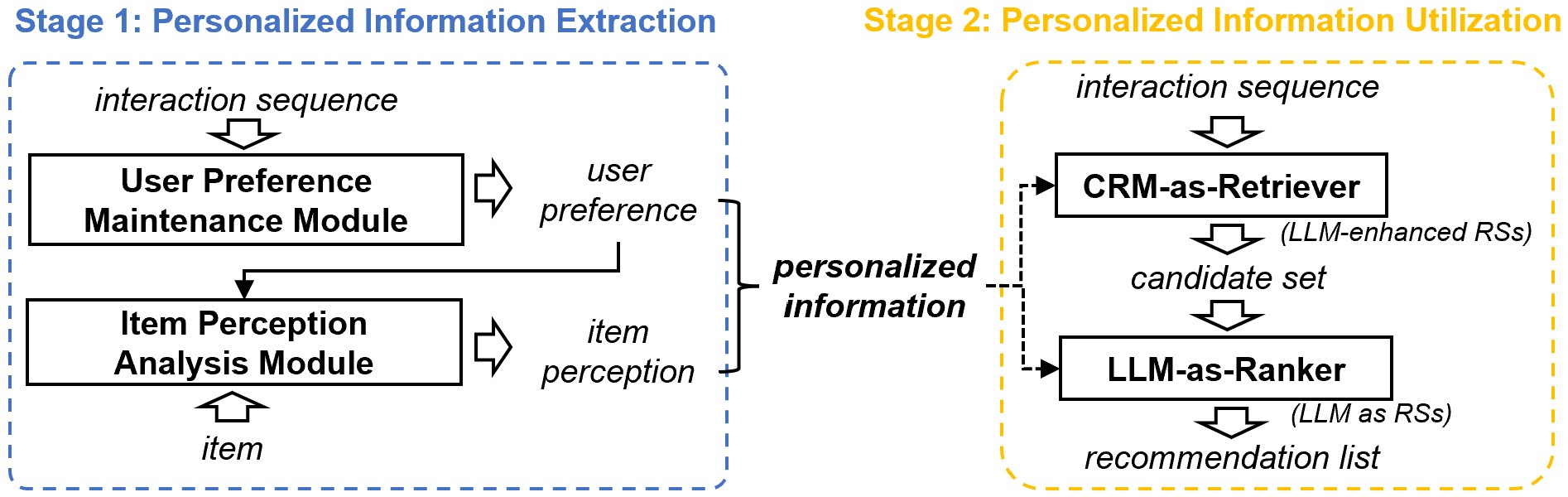}
  \caption{Overview of CoT-Rec.}
  \label{fig:nvoo}
\end{figure}
Due to their powerful understanding, reasoning, and generation capabilities, large language models (LLMs) have demonstrated remarkable success across multiple fields~\cite{sun2024adaptive,zhang2024agentcf,lu2024aligning,xia2022fire}.
Consequently, LLM-powered recommendations have attracted increasing attention~\cite{carraro2024enhancing,li2024explainable,zhang2023chatgpt,jiang2024item,liu2023personalized}.
LLMs contribute to recommender systems (RSs) in two ways:
(1) \textbf{LLMs as RSs}~\cite{zhang2024large,lei2024recexplainer,luo2024unlocking,liu2023autoseqrec}, where LLMs directly performing recommendation tasks, including \textit{LLM-as-Retriever} and \textit{LLM-as-Ranker};
(2) \textbf{LLM-enhanced RSs}~\cite{liu2024largesur,zhao2024let,liu2023recommendation}, where LLMs augment conventional recommendation models (CRMs).
\textit{Chain-of-Thought (CoT)}~\cite{wei2022chain} enhances LLMs' capacity for solving complex tasks by breaking down solutions into intermediate steps.
However, existing LLM-powered recommendation approaches often rely on LLMs to perform or enhance recommendation tasks directly based on users' interaction history, without explicitly modeling the reasoning process.
As a result, they do not fully exploit the reasoning capabilities of LLMs.

In this paper, we identify two key CoT processes---user preference analysis and item perception analysis---when performing or enhancing recommendation tasks with LLMs.
Furthermore, we propose a pipeline named \textbf{CoT-Rec} to incorporate user preference and item perception into LLM-powered recommendations.
As shown in Figure~\ref{fig:nvoo}, CoT-Rec consists of two stages: personalized information extraction and personalized information utilization.
In the \textbf{personalized information extraction} stage, a user preference maintenance module, inspired by the recurrent neural network (RNN)~\cite{medsker2001recurrent} architecture, analyzes the user's interaction sequence in chronological order and continuously updates user preference.
Subsequently, an item perception analysis module simulates the user's decision-making process through role-playing based on their preferences, thereby obtaining both the subjective perception and objective description of the item.
In the \textbf{personalized information utilization} stage, the extracted personalized information is integrated into the \textit{CRM-as-Retriever} to retrieve a candidate set from the entire item set.
This personalized information is then further fed into the \textit{LLM-as-Ranker} to optimize the ranking of the candidate set and generate the final recommendation list.
By explicitly incorporating user preferences and item perception, CoT-Rec enables LLM-powered recommendations to better leverage the reasoning capabilities of LLMs. 
Finally, experimental results on three datasets demonstrate that CoT-Rec improves the retrieval accuracy of the CRM in the retrieval stage and reduces the position bias of the LLM in the ranking stage.

\section{Related Work}

\subsection{LLM-as-RSs}

\subsubsection{LLM-as-Retriever}
LLM-as-Retriever leverages LLMs to recall a set of potentially relevant items from an entire item set based on a user's interaction history. 
\textbf{Bi-Step Grounding}~\cite{lin2024bridging,gao2024sprec,bao2023bi} retrieves items by measuring the similarity between the textual output of the LLM and the candidate set. 
\textbf{Indexing}~\cite{chen2024enhancing,li2024semantic,zheng2024adapting} discretizes items into semantically meaningful tokens and employs beam search for retrieval. 
\textbf{Modal Alignment}~\cite{yu2024break,li2023e4srec,chen2024hllm,zhang2024recgpt} transforms the semantic vectors encoded by the CRM to align them with the semantic space of the LLM, replacing the traditional next-token prediction head with a next-item prediction head. 

\subsubsection{LLM-as-Ranker}
The LLM-as-Ranker paradigm~\cite{cao2024aligning,luo2024recranker} requires LLMs to either rank a set of candidates based on a user's interaction history (list-wise ranking) or predict the likelihood of user interaction with a specific item (point-wise ranking).
Point-wise ranking predicts whether a specific item will be liked by a user~\cite{bao2023tallrec,lin2024clickprompt,lin2024rella,zheng2024harnessing,zhang2024text}, whereas list-wise ranking generates a personalized ranking for an entire set of candidate items~\cite{hou2024large,petruzzelli2024instructing,sun2024large,kim2024large,yue2023llamarec,liao2024llara,chen2024softmax,liu2022parameter,wang2024whole}.

\begin{figure}[t]
  \centering
  \includegraphics[width=0.99\linewidth]{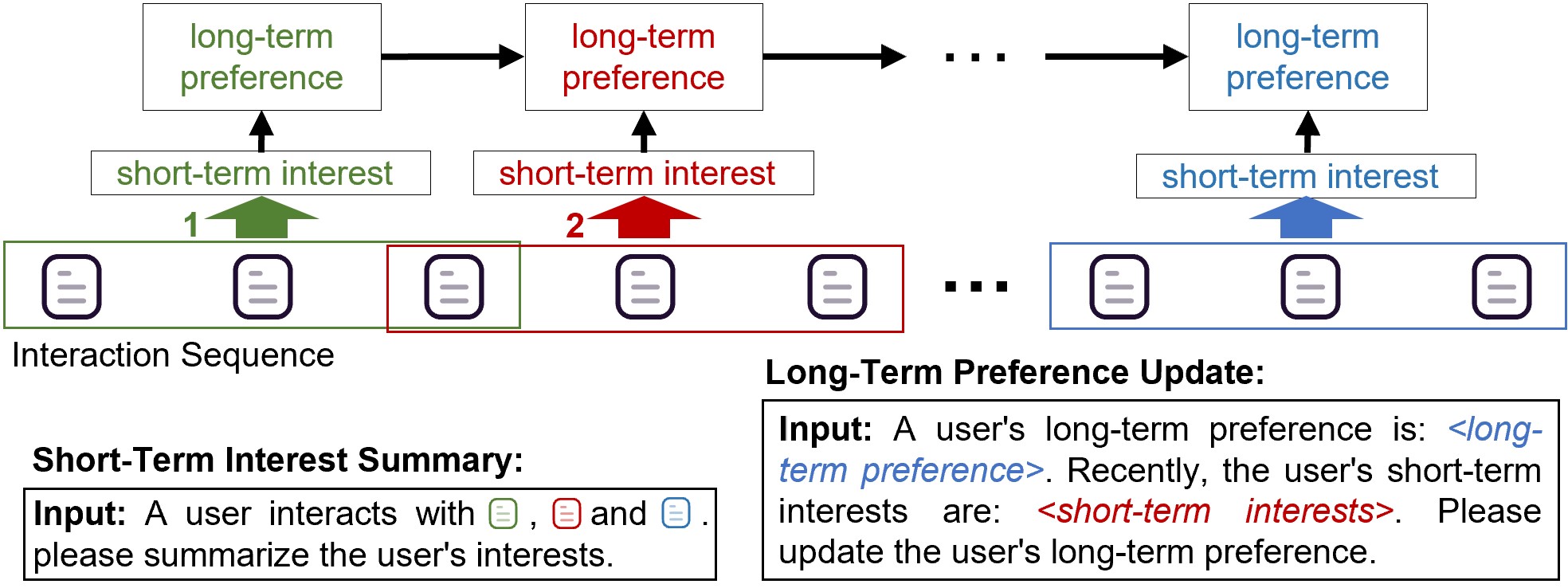}
  \caption{User preference maintenance module.}
  \label{fig:8mke}
\end{figure}

\subsection{LLM-enhanced RSs}
LLM-enhanced recommender systems (RSs) leverage LLMs to enhance the capabilities of CRM during the training phase, while LLMs are not required during inference. Depending on the type of knowledge provided by the LLM, some studies utilize LLMs to construct or optimize graphs that encode structural knowledge for CRM~\cite{hu2024bridging, zhang2024finerec, sakurai2024llm, wang2024llmrg, yang2024sequential,liu2024filtering}. Others introduce interaction information into CRM by generating synthetic interactions~\cite{wang2024large, wei2024llmrec}. Additionally, certain works enhance CRM inputs by optimizing features~\cite{jia2024altfs, wang2024llms,liu2023triple} or generating textual content~\cite{zhang2024embsum, du2024enhancing, sun2024largecf, xi2024towards}. Furthermore, some approaches improve CRM’s ability to learn high-quality representations by leveraging embeddings~\cite{geng2024breaking, wang2024can, cui2024distillation, liu2024large, harte2023leveraging, zhang2024notellm, ren2024representation}.

\section{Preliminaries}
SASRec and LlamaRec are the backbones used in this paper.


\textbf{SASRec}~\cite{kang2018self} consists of an input layer, an embedding layer, a sequence modeling layer, and a prediction layer. The input layer processes a sequence of item IDs, which are mapped to embeddings in the embedding layer. The sequence modeling layer employs a unidirectional Transformer to capture feature interactions and aggregate information across items. Finally, the prediction layer outputs a probability distribution over the next item.

\begin{figure}[t]
  \centering
  \includegraphics[width=0.9\linewidth]{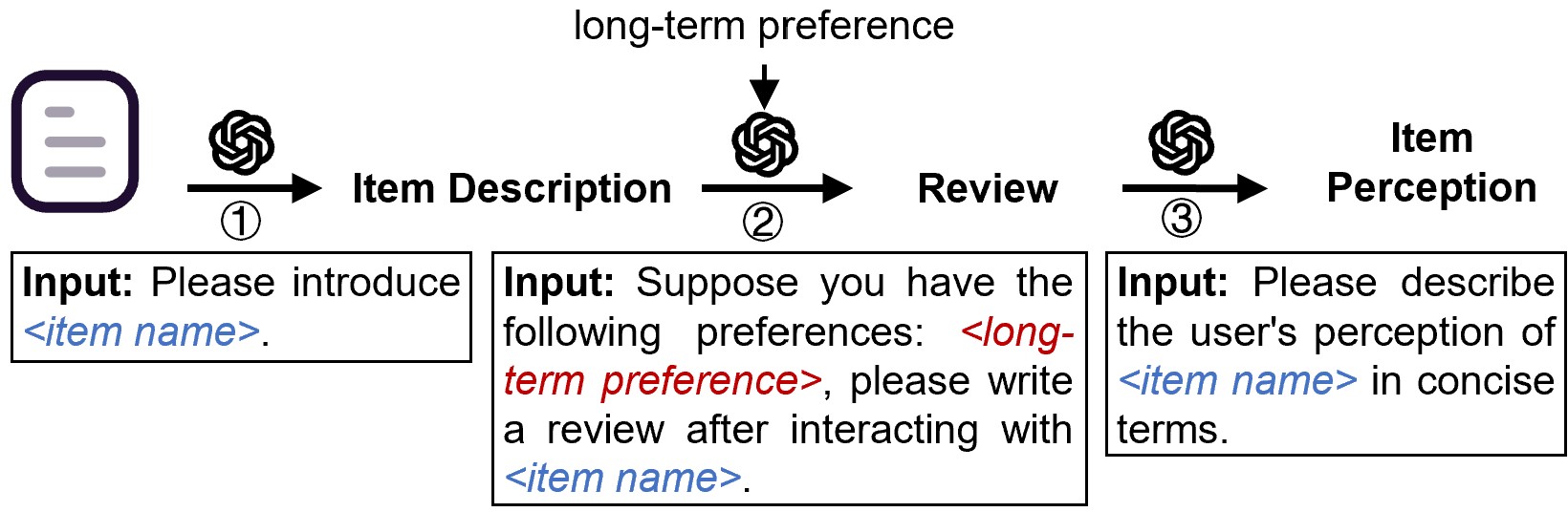}
  \caption{Item perception analysis module.}
  \label{fig:pc9l}
\end{figure}

\textbf{LlamaRec}~\cite{yue2023llamarec} follows an LLM-as-Ranker approach. It first fine-tunes the LLM, which, given a candidate set, is instructed to predict the index (e.g., A, B, C, D) of the item the user is most likely to engage with. Based on the distribution of the predicted indices, the candidate set is then ranked in a list-wise manner.

\begin{figure*}[t]
  \centering
  \includegraphics[width=0.9\linewidth]{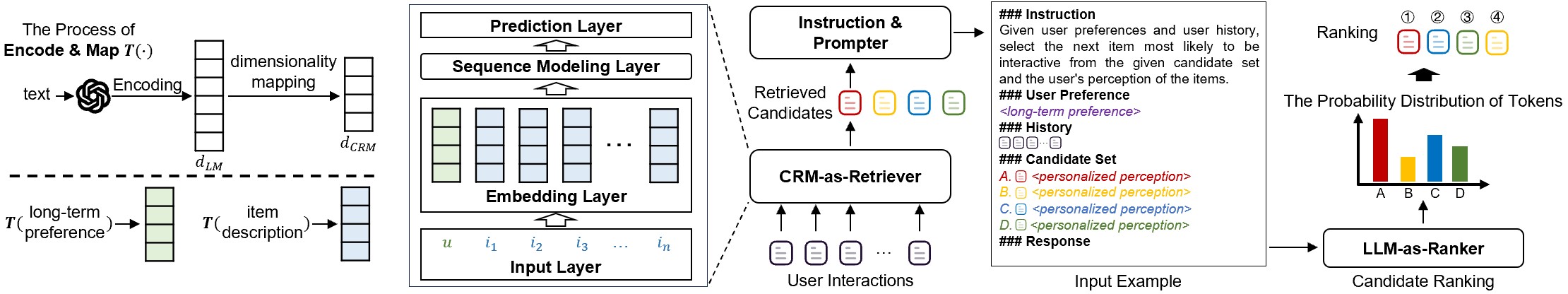}
  \caption{Model architecture of the personalized information utilization stage in CoT-Rec. Note that we have added user embedding as an input for the CRM and included extracted user preferences and subjective perception of items in the prompt.}
  \label{fig:9cn6}
\end{figure*}

\section{Methods}
We have introduced the overall process of CoT-Rec through Figure~\ref{fig:nvoo}. Here, we present its details.

\subsection{Personalized Information Extraction}
The personalized information extraction stage extracts user preferences and item perception based on a user's interaction sequence.

\subsubsection{User Preference Maintenance Module}
As shown in Figure \ref{fig:8mke}, this module extracts user preferences by analyzing interaction sequences. Prior research suggests that even when temporal information is explicitly highlighted in the prompt, LLMs struggle to capture temporal relationships within interaction sequences accurately~\cite{hou2024large}. To improve LLMs' temporal sensitivity, we take inspiration from RNNs and process items in batches following chronological order. Each batch's processing result is summarized as short-term interest, which helps derive long-term preferences. Specifically, the short-term interest from the first batch directly serves as the initial long-term preference. Thereafter, each new short-term interest is integrated with the preceding long-term preference to update it. Additionally, adjacent batches partially overlap to maintain information continuity.

\subsubsection{Item Perception Analysis Module}
As shown in Figure \ref{fig:pc9l}, this module enables the LLM to capture a user's perception of a specific item through role-playing. The process consists of three steps. First, the LLM generates an objective description of the item. Next, based on the extracted long-term preference, the LLM simulates the user writing a review of the item. Finally, key terms are extracted from the review to summarize the user's subjective impression.
By adopting a role-playing approach, the LLM generates differentiated perception information based on varying user preferences, thereby capturing the diversity in individual understandings of the same item.
Notably, both objective descriptions and subjective perceptions can be utilized.

\subsubsection{Scalability}
All operations in the personalized information extraction stage are performed offline, ensuring no impact on inference efficiency during the service phase.

\subsection{Personalized Information Utilization}
Figure \ref{fig:9cn6} shows the model architecture of the personalized information utilization stage in CoT-Rec.

\subsubsection{Architecture}
The retrieval stage demands very high throughput. However, LLMs’ inference speed is several orders of magnitude lower than that of CRMs. As a result, employing LLMs as retrieval models is computationally prohibitive and fails to meet the requirements of large-scale users. Therefore, we argue that CRMs are more suitable for the retrieval stage. While LLMs also face computational efficiency challenges in ranking tasks, studies suggest that ranking can be deployed on the client side~\cite{liu2024filtering}. Since ranking involves only reordering the candidate set retrieved by the server rather than processing the entire item corpus, the LLM-as-Ranker approach could be executed on the client side, mitigating its impact on server-side inference efficiency.

In summary, we adopt CRM-as-Retriever and LLM-as-Ranker, to generate the final recommendation list. In this framework, CRM-as-Retriever serves as the retrieval model, which recalls a candidate set from the complete item set. Meanwhile, LLM-as-Ranker ranks the retrieved candidates to produce the final recommendations. Our approach is model-agnostic. In this paper, we introduce it using SASRec as the CRM and LlamaRec as the LLM.

\subsubsection{CRM-as-Retriever}
We define the \textit{Encode \& Map} operation as $T: \text{string} \rightarrow \mathbb{R}^{d_\text{LM}} \rightarrow \mathbb{R}^{d_\text{CRM}}$. Specifically, a language model (LM) first encodes the text into an embedding of dimension $d_\text{LM}$, which is then mapped into the CRM embedding space of dimension $d_\text{CRM}$ via a dimensionality reduction technique.

Previous research suggests that initializing the embedding layer of SASRec with item captions processed through the Encode \& Map operation can improve recommendation accuracy. However, this approach has two main limitations: (1) item captions provide limited information; and (2) SASRec relies exclusively on item sequences, omitting user information.

To address these limitations, we enhance SASRec by prepending the user ID to the item interaction sequence, explicitly linking it to a specific user. We further apply the Encode \& Map operation to user preferences, using the resulting embeddings to initialize user representations. Similarly, we process both item captions and descriptions through the Encode \& Map operation to initialize item embeddings. This approach ensures that CRM inputs incorporate user preferences and item descriptions---two essential intermediate results in the CoT process.

\subsubsection{LLM-as-Ranker}
Building on LlamaRec, we integrate user preferences and users' perceptions of candidate items, obtained during the personalized information extraction phase, into the prompt. Additionally, we construct an instruction-tuning dataset incorporating personalized information and leverage LoRA~\cite{hu2021lora} to fine-tune the LLM for ranking tasks.

\subsubsection{Scalability}
In the retrieval stage, the improved CRM differs from the original only by incorporating an additional user embedding as input. Consequently, the optimization in CRM-as-Retriever has minimal impact on inference efficiency. Furthermore, LLM-as-Ranker is deployed on the user side, ensuring that its inference process imposes no additional computational burden on the server.

\section{Experiments}
In this section, we empirically evaluate whether CoT-Rec improves the performance of CRM-as-Retriever and LLM-as-Ranker. We do not compare it with other LLM-powered recommendation methods, as our focus is on exploring how CoT-Rec can effectively integrate with existing methods to achieve greater performance gains, rather than conducting a comprehensive comparison of different recommendation architectures. In fact, many architectures can leverage CoT-Rec to enhance their performance.

\subsection{Settings}

\subsubsection{Datasets}
We conducted experiments on three datasets from distinct domains: Amazon Review (Food)~\cite{hou2024bridging} (e-commerce), MIND~\cite{wu2020mind} (news), and Yelp (reviews). For each dataset, we retained users and items with at least five interactions and ordered them chronologically.

\subsubsection{Evaluation}
For CRM-as-Retriever, the task is to retrieve the correct item from the entire set. We evaluate CRM-as-Retriever using the leave-one-out method, which is widely adopted for assessing sequential recommendation methods. The accuracy of the retrieval stage is measured using Hit@K and NDCG@K.
For LLM-as-Ranker, the task is to rank the candidate set retrieved in the previous stage. We evaluate the accuracy and position bias of the ranking stage using NDCG@K and MAPB. Position bias refers to the effect where the LLM’s output is influenced by the target item's position within the candidate set. We fix $K=10$ and report the average over five independent runs.

\textbf{Mean Absolute Position Bias (MAPB).} We propose Mean Absolute Position Bias (MAPB) to quantify the position bias in LLMs. Assuming the candidate set size is $M$, the sample bias is defined as: $\text{Sample Bias}_i = \frac{1}{M} \sum_{j=1}^M |r_{i,j} - \bar{r}_i|\text{,}$
where $r_{i,j}$ is the predicted rank of the target item when placed in position $j$, and $\bar{r}_i = \frac{1}{M} \sum_{j=1}^M r_{i,j}$ is the average predicted rank of the target item for sample $i$. Then, MAPB is defined as: $\text{MAPB} = \frac{1}{n} \sum_{i=1}^n \text{Sample Bias}_i\text{,}$
where $n$ is the number of the samples. MAPB measures the average absolute deviation of the target item's predicted rank from its mean rank across all possible positions and all samples. A lower MAPB indicates that the model is less sensitive to the target item's position, which suggests better robustness to position bias.

\subsubsection{Backbones}
We select SASRec as the CRM for the retrieval stage, and Qwen2.5-7B-Instruct as the LLM for the ranking stage.

\begin{table}[]\footnotesize
\caption{Results of CRM-as-Retriever.}
\label{tab:exp1}
\begin{tabular}{c|c|cc|cc|cc}
\hline
\multirow{2}{*}{\textbf{\begin{tabular}[c]{@{}c@{}}User\\ Embedding\end{tabular}}} & \multirow{2}{*}{\textbf{\begin{tabular}[c]{@{}c@{}}Item\\ Embedding\end{tabular}}} & \multicolumn{2}{c|}{\textbf{MIND}} & \multicolumn{2}{c|}{\textbf{Food}} & \multicolumn{2}{c}{\textbf{Yelp}} \\ \cline{3-8} 
                                                                                   &                                                                                    & \textbf{Hit}    & \textbf{NDCG}    & \textbf{Hit}    & \textbf{NDCG}    & \textbf{Hit}    & \textbf{NDCG}   \\ \hline
\multirow{3}{*}{None}                                                              & Random                                                                             & 0.1387          & 0.0737           & 0.0182          & 0.0096           & 0.0297          & 0.0148          \\
                                                                                   & Caption                                                                            & 0.1447          & 0.0776           & 0.0263          & 0.0141           & 0.0306          & 0.0153          \\
                                                                                   & Description                                                                            & 0.1498          & 0.0807           & 0.0281          & 0.0150           & 0.0313          & 0.0155          \\ \hline
\multirow{2}{*}{Random}                                                            & Caption                                                                            & 0.1382          & 0.0740           & 0.0275          & 0.0148           & 0.0305          & 0.0153          \\
                                                                                   & Description                                                                            & 0.1396          & 0.0747           & 0.0284          & 0.0155           & 0.0313          & 0.0157          \\ \hline
\multirow{2}{*}{Preference}                                                        & Caption                                                                            & 0.1539          & 0.0831           & 0.0291          & 0.0156           & 0.0346          & 0.0170          \\
                                                                                   & Description                                                                            & 0.1551          & 0.0836           & 0.0295          & 0.0158           & 0.0348          & 0.0171          \\ \hline
\end{tabular}
\end{table}

\subsection{Results}

\subsubsection{CRM-as-Retriever}
Table \ref{tab:exp1} presents the performance of CRM-as-Retriever under different enhancement methods. When User Embedding is set to ``None'', it means no user embedding is used; ``Random'' indicates random initialization, while ``Preference'' refers to initialization using the Encode \& Map operation based on user preferences. Item embeddings are initialized using one of the following methods: ``Random'' (random initialization), ``Caption'' (initialized via Encode \& Map using item captions), and ``Description'' (initialized via Encode \& Map using both item captions and descriptions). 

When User Embedding is set to ``None'', item embedding performance follows the order: ``Description > Caption > Random''. This indicates that incorporating item information improves CRM performance, and descriptions provide richer information than captions. However, when User Embedding is set to ``Random'', performance does not always surpass that of ``None''. This suggests that introducing user embedding does not necessarily enhance performance, which aligns with SASRec’s conclusion that additional user embedding does not improve performance~\cite{kang2018self}. However, when User Embedding is set to ``Preference'', performance consistently surpasses that of ``None'', demonstrating that user preference-based embedding initialization improves model effectiveness.

\subsubsection{LLM-as-Ranker}
Table \ref{tab:exp2} presents the effectiveness of different candidate set ranking methods.
Here, Retriever refers to the retrieval model used. ``CRM'' refers to the basic SASRec model (as in Table \ref{tab:exp1} with ``User Embedding=None'' and ``Item Embedding=Random''), while ``CRM++'' denotes its improved version (as in Table \ref{tab:exp1} with ``User Embedding=Preference'' and ``Item Embedding=Description'').
The Ranker denotes the ranking method. ``None'' indicates that no LLM-based ranking is applied, and the CRM retrieval results are used directly. ``LLM'' performs ranking without user preference and item perception. 
``LLM++'' incorporates user preferences and subjective item perception.

The experimental results show that ranking with ``LLM'' and ``LLM++'' generally leads to a significant improvement in recommendation performance compared to no ranking (``Ranker=None'').
Additionally, while ``LLM++'' offers only a slight improvement in accuracy over ``LLM'', it substantially reduce position bias. 
This may be because integrating user preferences and personalized perception helps the LLM focus on task-relevant information, thereby mitigating its inherent position bias in ranking.

\begin{table}[]\footnotesize
\caption{Results of LLM-as-Ranker.}
\label{tab:exp2}
\begin{tabular}{c|c|cc|cc|cc}
\hline
\multirow{2}{*}{\textbf{Retriever}} & \multirow{2}{*}{\textbf{Ranker}} & \multicolumn{2}{c|}{\textbf{MIND}} & \multicolumn{2}{c|}{\textbf{Food}} & \multicolumn{2}{c}{\textbf{Yelp}} \\ \cline{3-8} 
                                    &                                  & \textbf{NDCG}    & \textbf{MAPB}   & \textbf{NDCG}    & \textbf{MAPB}   & \textbf{NDCG}   & \textbf{MAPB}   \\ \hline
\multirow{3}{*}{CRM}             & None                                & 0.0737           & -               & 0.0096           & -               & 0.0148          & -               \\
                                    & LLM                              & 0.0763           & 0.9723          & 0.0112           & 0.9819          & 0.0148          & 1.4928          \\
                                    & LLM++                            & 0.0765           & 0.8413          & 0.0113           & 0.8086          & 0.0151          & 1.1060          \\ \hline
\multirow{3}{*}{CRM++}           & None                                & 0.0836           & -               & 0.0158           & -               & 0.0171          & -               \\
                                    & LLM                              & 0.0837           & 1.0336          & 0.0176           & 1.0664          & 0.0172          & 1.3716          \\
                                    & LLM++                            & 0.0840           & 0.8168          & 0.0177           & 0.7641          & 0.0174          & 1.1964          \\ \hline
\end{tabular}
\end{table}

\section{Conclusions}
We propose CoT-Rec, which enhances LLM-powered recommendations via two key CoT processes: user preference analysis and item perception analysis. Experimental results effectively validate our approach. Inference efficiency analysis shows that CoT-Rec introduces negligible overhead, suggesting its potential for industrial applications.

\begin{acks}
This work is supported by National Natural Science Foundation of China (NSFC) under the Grant No. 62372113. Peng Zhang is a faculty of School of Computer Science, Fudan University. Tun Lu is a faculty of School of Computer Science, Shanghai Key Laboratory of Data Science, Fudan Institute on Aging, MOE Laboratory for National Development and Intelligent Governance, and Shanghai Institute of Intelligent Electronics \& Systems, Fudan University.
\end{acks}

\bibliographystyle{ACM-Reference-Format}
\balance
\bibliography{main}


\begin{thebibliography}{70}


\ifx \showCODEN    \undefined \def \showCODEN     #1{\unskip}     \fi
\ifx \showISBNx    \undefined \def \showISBNx     #1{\unskip}     \fi
\ifx \showISBNxiii \undefined \def \showISBNxiii  #1{\unskip}     \fi
\ifx \showISSN     \undefined \def \showISSN      #1{\unskip}     \fi
\ifx \showLCCN     \undefined \def \showLCCN      #1{\unskip}     \fi
\ifx \shownote     \undefined \def \shownote      #1{#1}          \fi
\ifx \showarticletitle \undefined \def \showarticletitle #1{#1}   \fi
\ifx \showURL      \undefined \def \showURL       {\relax}        \fi
\providecommand\bibfield[2]{#2}
\providecommand\bibinfo[2]{#2}
\providecommand\natexlab[1]{#1}
\providecommand\showeprint[2][]{arXiv:#2}

\bibitem[Bao et~al\mbox{.}(2023a)]%
        {bao2023bi}
\bibfield{author}{\bibinfo{person}{Keqin Bao}, \bibinfo{person}{Jizhi Zhang}, \bibinfo{person}{Wenjie Wang}, \bibinfo{person}{Yang Zhang}, \bibinfo{person}{Zhengyi Yang}, \bibinfo{person}{Yancheng Luo}, \bibinfo{person}{Chong Chen}, \bibinfo{person}{Fuli Feng}, {and} \bibinfo{person}{Qi Tian}.} \bibinfo{year}{2023}\natexlab{a}.
\newblock \showarticletitle{A bi-step grounding paradigm for large language models in recommendation systems}.
\newblock \bibinfo{journal}{\emph{ACM Transactions on Recommender Systems}} (\bibinfo{year}{2023}).
\newblock


\bibitem[Bao et~al\mbox{.}(2023b)]%
        {bao2023tallrec}
\bibfield{author}{\bibinfo{person}{Keqin Bao}, \bibinfo{person}{Jizhi Zhang}, \bibinfo{person}{Yang Zhang}, \bibinfo{person}{Wenjie Wang}, \bibinfo{person}{Fuli Feng}, {and} \bibinfo{person}{Xiangnan He}.} \bibinfo{year}{2023}\natexlab{b}.
\newblock \showarticletitle{Tallrec: An effective and efficient tuning framework to align large language model with recommendation}. In \bibinfo{booktitle}{\emph{Proceedings of the 17th ACM Conference on Recommender Systems}}. \bibinfo{pages}{1007--1014}.
\newblock


\bibitem[Cao et~al\mbox{.}(2024)]%
        {cao2024aligning}
\bibfield{author}{\bibinfo{person}{Yuwei Cao}, \bibinfo{person}{Nikhil Mehta}, \bibinfo{person}{Xinyang Yi}, \bibinfo{person}{Raghunandan Keshavan}, \bibinfo{person}{Lukasz Heldt}, \bibinfo{person}{Lichan Hong}, \bibinfo{person}{Ed~H Chi}, {and} \bibinfo{person}{Maheswaran Sathiamoorthy}.} \bibinfo{year}{2024}\natexlab{}.
\newblock \showarticletitle{Aligning Large Language Models with Recommendation Knowledge}. In \bibinfo{booktitle}{\emph{Findings of the Association for Computational Linguistics: NAACL 2024}}. \bibinfo{pages}{1051--1066}.
\newblock


\bibitem[Carraro and Bridge(2024)]%
        {carraro2024enhancing}
\bibfield{author}{\bibinfo{person}{Diego Carraro} {and} \bibinfo{person}{Derek Bridge}.} \bibinfo{year}{2024}\natexlab{}.
\newblock \showarticletitle{Enhancing recommendation diversity by re-ranking with large language models}.
\newblock \bibinfo{journal}{\emph{ACM Transactions on Recommender Systems}} (\bibinfo{year}{2024}).
\newblock


\bibitem[Chen et~al\mbox{.}(2024a)]%
        {chen2024hllm}
\bibfield{author}{\bibinfo{person}{Junyi Chen}, \bibinfo{person}{Lu Chi}, \bibinfo{person}{Bingyue Peng}, {and} \bibinfo{person}{Zehuan Yuan}.} \bibinfo{year}{2024}\natexlab{a}.
\newblock \showarticletitle{Hllm: Enhancing sequential recommendations via hierarchical large language models for item and user modeling}.
\newblock \bibinfo{journal}{\emph{arXiv preprint arXiv:2409.12740}} (\bibinfo{year}{2024}).
\newblock


\bibitem[Chen et~al\mbox{.}(2024b)]%
        {chen2024enhancing}
\bibfield{author}{\bibinfo{person}{Runjin Chen}, \bibinfo{person}{Mingxuan Ju}, \bibinfo{person}{Ngoc Bui}, \bibinfo{person}{Dimosthenis Antypas}, \bibinfo{person}{Stanley Cai}, \bibinfo{person}{Xiaopeng Wu}, \bibinfo{person}{Leonardo Neves}, \bibinfo{person}{Zhangyang Wang}, \bibinfo{person}{Neil Shah}, {and} \bibinfo{person}{Tong Zhao}.} \bibinfo{year}{2024}\natexlab{b}.
\newblock \showarticletitle{Enhancing Item Tokenization for Generative Recommendation through Self-Improvement}.
\newblock \bibinfo{journal}{\emph{arXiv preprint arXiv:2412.17171}} (\bibinfo{year}{2024}).
\newblock


\bibitem[Chen et~al\mbox{.}(2024c)]%
        {chen2024softmax}
\bibfield{author}{\bibinfo{person}{Yuxin Chen}, \bibinfo{person}{Junfei Tan}, \bibinfo{person}{An Zhang}, \bibinfo{person}{Zhengyi Yang}, \bibinfo{person}{Leheng Sheng}, \bibinfo{person}{Enzhi Zhang}, \bibinfo{person}{Xiang Wang}, {and} \bibinfo{person}{Tat-Seng Chua}.} \bibinfo{year}{2024}\natexlab{c}.
\newblock \showarticletitle{On Softmax Direct Preference Optimization for Recommendation}. In \bibinfo{booktitle}{\emph{Advances in Neural Information Processing Systems}}. \bibinfo{pages}{27463--27489}.
\newblock


\bibitem[Cui et~al\mbox{.}(2024)]%
        {cui2024distillation}
\bibfield{author}{\bibinfo{person}{Yu Cui}, \bibinfo{person}{Feng Liu}, \bibinfo{person}{Pengbo Wang}, \bibinfo{person}{Bohao Wang}, \bibinfo{person}{Heng Tang}, \bibinfo{person}{Yi Wan}, \bibinfo{person}{Jun Wang}, {and} \bibinfo{person}{Jiawei Chen}.} \bibinfo{year}{2024}\natexlab{}.
\newblock \showarticletitle{Distillation matters: empowering sequential recommenders to match the performance of large language models}. In \bibinfo{booktitle}{\emph{Proceedings of the 18th ACM Conference on Recommender Systems}}. \bibinfo{pages}{507--517}.
\newblock


\bibitem[Du et~al\mbox{.}(2024)]%
        {du2024enhancing}
\bibfield{author}{\bibinfo{person}{Yingpeng Du}, \bibinfo{person}{Di Luo}, \bibinfo{person}{Rui Yan}, \bibinfo{person}{Xiaopei Wang}, \bibinfo{person}{Hongzhi Liu}, \bibinfo{person}{Hengshu Zhu}, \bibinfo{person}{Yang Song}, {and} \bibinfo{person}{Jie Zhang}.} \bibinfo{year}{2024}\natexlab{}.
\newblock \showarticletitle{Enhancing job recommendation through llm-based generative adversarial networks}. In \bibinfo{booktitle}{\emph{Proceedings of the AAAI Conference on Artificial Intelligence}}, Vol.~\bibinfo{volume}{38}. \bibinfo{pages}{8363--8371}.
\newblock


\bibitem[Gao et~al\mbox{.}(2024)]%
        {gao2024sprec}
\bibfield{author}{\bibinfo{person}{Chongming Gao}, \bibinfo{person}{Ruijun Chen}, \bibinfo{person}{Shuai Yuan}, \bibinfo{person}{Kexin Huang}, \bibinfo{person}{Yuanqing Yu}, {and} \bibinfo{person}{Xiangnan He}.} \bibinfo{year}{2024}\natexlab{}.
\newblock \showarticletitle{SPRec: Leveraging Self-Play to Debias Preference Alignment for Large Language Model-based Recommendations}.
\newblock \bibinfo{journal}{\emph{arXiv preprint arXiv:2412.09243}} (\bibinfo{year}{2024}).
\newblock


\bibitem[Geng et~al\mbox{.}(2024)]%
        {geng2024breaking}
\bibfield{author}{\bibinfo{person}{Binzong Geng}, \bibinfo{person}{Zhaoxin Huan}, \bibinfo{person}{Xiaolu Zhang}, \bibinfo{person}{Yong He}, \bibinfo{person}{Liang Zhang}, \bibinfo{person}{Fajie Yuan}, \bibinfo{person}{Jun Zhou}, {and} \bibinfo{person}{Linjian Mo}.} \bibinfo{year}{2024}\natexlab{}.
\newblock \showarticletitle{Breaking the length barrier: Llm-enhanced CTR prediction in long textual user behaviors}. In \bibinfo{booktitle}{\emph{Proceedings of the 47th International ACM SIGIR Conference on Research and Development in Information Retrieval}}. \bibinfo{pages}{2311--2315}.
\newblock


\bibitem[Harte et~al\mbox{.}(2023)]%
        {harte2023leveraging}
\bibfield{author}{\bibinfo{person}{Jesse Harte}, \bibinfo{person}{Wouter Zorgdrager}, \bibinfo{person}{Panos Louridas}, \bibinfo{person}{Asterios Katsifodimos}, \bibinfo{person}{Dietmar Jannach}, {and} \bibinfo{person}{Marios Fragkoulis}.} \bibinfo{year}{2023}\natexlab{}.
\newblock \showarticletitle{Leveraging large language models for sequential recommendation}. In \bibinfo{booktitle}{\emph{Proceedings of the 17th ACM Conference on Recommender Systems}}. \bibinfo{pages}{1096--1102}.
\newblock


\bibitem[Hou et~al\mbox{.}(2024a)]%
        {hou2024bridging}
\bibfield{author}{\bibinfo{person}{Yupeng Hou}, \bibinfo{person}{Jiacheng Li}, \bibinfo{person}{Zhankui He}, \bibinfo{person}{An Yan}, \bibinfo{person}{Xiusi Chen}, {and} \bibinfo{person}{Julian McAuley}.} \bibinfo{year}{2024}\natexlab{a}.
\newblock \showarticletitle{Bridging Language and Items for Retrieval and Recommendation}.
\newblock \bibinfo{journal}{\emph{arXiv preprint arXiv:2403.03952}} (\bibinfo{year}{2024}).
\newblock


\bibitem[Hou et~al\mbox{.}(2024b)]%
        {hou2024large}
\bibfield{author}{\bibinfo{person}{Yupeng Hou}, \bibinfo{person}{Junjie Zhang}, \bibinfo{person}{Zihan Lin}, \bibinfo{person}{Hongyu Lu}, \bibinfo{person}{Ruobing Xie}, \bibinfo{person}{Julian McAuley}, {and} \bibinfo{person}{Wayne~Xin Zhao}.} \bibinfo{year}{2024}\natexlab{b}.
\newblock \showarticletitle{Large language models are zero-shot rankers for recommender systems}. In \bibinfo{booktitle}{\emph{European Conference on Information Retrieval}}. Springer, \bibinfo{pages}{364--381}.
\newblock


\bibitem[Hu et~al\mbox{.}(2021)]%
        {hu2021lora}
\bibfield{author}{\bibinfo{person}{Edward~J Hu}, \bibinfo{person}{Yelong Shen}, \bibinfo{person}{Phillip Wallis}, \bibinfo{person}{Zeyuan Allen-Zhu}, \bibinfo{person}{Yuanzhi Li}, \bibinfo{person}{Shean Wang}, \bibinfo{person}{Lu Wang}, {and} \bibinfo{person}{Weizhu Chen}.} \bibinfo{year}{2021}\natexlab{}.
\newblock \showarticletitle{Lora: Low-rank adaptation of large language models}.
\newblock \bibinfo{journal}{\emph{arXiv preprint arXiv:2106.09685}} (\bibinfo{year}{2021}).
\newblock


\bibitem[Hu et~al\mbox{.}(2024)]%
        {hu2024bridging}
\bibfield{author}{\bibinfo{person}{Zheng Hu}, \bibinfo{person}{Zhe Li}, \bibinfo{person}{Ziyun Jiao}, \bibinfo{person}{Satoshi Nakagawa}, \bibinfo{person}{Jiawen Deng}, \bibinfo{person}{Shimin Cai}, \bibinfo{person}{Tao Zhou}, {and} \bibinfo{person}{Fuji Ren}.} \bibinfo{year}{2024}\natexlab{}.
\newblock \showarticletitle{Bridging the User-side Knowledge Gap in Knowledge-aware Recommendations with Large Language Models}.
\newblock \bibinfo{journal}{\emph{arXiv preprint arXiv:2412.13544}} (\bibinfo{year}{2024}).
\newblock


\bibitem[Jia et~al\mbox{.}(2024)]%
        {jia2024altfs}
\bibfield{author}{\bibinfo{person}{Pengyue Jia}, \bibinfo{person}{Zhaocheng Du}, \bibinfo{person}{Yichao Wang}, \bibinfo{person}{Xiangyu Zhao}, \bibinfo{person}{Xiaopeng Li}, \bibinfo{person}{Yuhao Wang}, \bibinfo{person}{Qidong Liu}, \bibinfo{person}{Huifeng Guo}, {and} \bibinfo{person}{Ruiming Tang}.} \bibinfo{year}{2024}\natexlab{}.
\newblock \showarticletitle{AltFS: Agency-light Feature Selection with Large Language Models in Deep Recommender Systems}.
\newblock \bibinfo{journal}{\emph{arXiv preprint arXiv:2412.08516}} (\bibinfo{year}{2024}).
\newblock


\bibitem[Jiang et~al\mbox{.}(2024)]%
        {jiang2024item}
\bibfield{author}{\bibinfo{person}{Meng Jiang}, \bibinfo{person}{Keqin Bao}, \bibinfo{person}{Jizhi Zhang}, \bibinfo{person}{Wenjie Wang}, \bibinfo{person}{Zhengyi Yang}, \bibinfo{person}{Fuli Feng}, {and} \bibinfo{person}{Xiangnan He}.} \bibinfo{year}{2024}\natexlab{}.
\newblock \showarticletitle{Item-side Fairness of Large Language Model-based Recommendation System}. In \bibinfo{booktitle}{\emph{Proceedings of the ACM on Web Conference 2024}}. \bibinfo{pages}{4717--4726}.
\newblock


\bibitem[Kang and McAuley(2018)]%
        {kang2018self}
\bibfield{author}{\bibinfo{person}{Wang-Cheng Kang} {and} \bibinfo{person}{Julian McAuley}.} \bibinfo{year}{2018}\natexlab{}.
\newblock \showarticletitle{Self-attentive sequential recommendation}. In \bibinfo{booktitle}{\emph{2018 IEEE international conference on data mining (ICDM)}}. IEEE, \bibinfo{pages}{197--206}.
\newblock


\bibitem[Kim et~al\mbox{.}(2024)]%
        {kim2024large}
\bibfield{author}{\bibinfo{person}{Sein Kim}, \bibinfo{person}{Hongseok Kang}, \bibinfo{person}{Seungyoon Choi}, \bibinfo{person}{Donghyun Kim}, \bibinfo{person}{Minchul Yang}, {and} \bibinfo{person}{Chanyoung Park}.} \bibinfo{year}{2024}\natexlab{}.
\newblock \showarticletitle{Large language models meet collaborative filtering: An efficient all-round llm-based recommender system}. In \bibinfo{booktitle}{\emph{Proceedings of the 30th ACM SIGKDD Conference on Knowledge Discovery and Data Mining}}. \bibinfo{pages}{1395--1406}.
\newblock


\bibitem[Lei et~al\mbox{.}(2024)]%
        {lei2024recexplainer}
\bibfield{author}{\bibinfo{person}{Yuxuan Lei}, \bibinfo{person}{Jianxun Lian}, \bibinfo{person}{Jing Yao}, \bibinfo{person}{Xu Huang}, \bibinfo{person}{Defu Lian}, {and} \bibinfo{person}{Xing Xie}.} \bibinfo{year}{2024}\natexlab{}.
\newblock \showarticletitle{Recexplainer: Aligning large language models for explaining recommendation models}. In \bibinfo{booktitle}{\emph{Proceedings of the 30th ACM SIGKDD Conference on Knowledge Discovery and Data Mining}}. \bibinfo{pages}{1530--1541}.
\newblock


\bibitem[Li et~al\mbox{.}(2024b)]%
        {li2024semantic}
\bibfield{author}{\bibinfo{person}{Guanghan Li}, \bibinfo{person}{Xun Zhang}, \bibinfo{person}{Yufei Zhang}, \bibinfo{person}{Yifan Yin}, \bibinfo{person}{Guojun Yin}, {and} \bibinfo{person}{Wei Lin}.} \bibinfo{year}{2024}\natexlab{b}.
\newblock \showarticletitle{Semantic Convergence: Harmonizing Recommender Systems via Two-Stage Alignment and Behavioral Semantic Tokenization}.
\newblock \bibinfo{journal}{\emph{arXiv preprint arXiv:2412.13771}} (\bibinfo{year}{2024}).
\newblock


\bibitem[Li et~al\mbox{.}(2023)]%
        {li2023e4srec}
\bibfield{author}{\bibinfo{person}{Xinhang Li}, \bibinfo{person}{Chong Chen}, \bibinfo{person}{Xiangyu Zhao}, \bibinfo{person}{Yong Zhang}, {and} \bibinfo{person}{Chunxiao Xing}.} \bibinfo{year}{2023}\natexlab{}.
\newblock \showarticletitle{E4srec: An elegant effective efficient extensible solution of large language models for sequential recommendation}.
\newblock \bibinfo{journal}{\emph{arXiv preprint arXiv:2312.02443}} (\bibinfo{year}{2023}).
\newblock


\bibitem[Li et~al\mbox{.}(2024a)]%
        {li2024explainable}
\bibfield{author}{\bibinfo{person}{Zelong Li}, \bibinfo{person}{Yan Liang}, \bibinfo{person}{Ming Wang}, \bibinfo{person}{Sungro Yoon}, \bibinfo{person}{Jiaying Shi}, \bibinfo{person}{Xin Shen}, \bibinfo{person}{Xiang He}, \bibinfo{person}{Chenwei Zhang}, \bibinfo{person}{Wenyi Wu}, \bibinfo{person}{Hanbo Wang}, {et~al\mbox{.}}} \bibinfo{year}{2024}\natexlab{a}.
\newblock \showarticletitle{Explainable and coherent complement recommendation based on large language models}. In \bibinfo{booktitle}{\emph{Proceedings of the 33rd ACM International Conference on Information and Knowledge Management}}. \bibinfo{pages}{4678--4685}.
\newblock


\bibitem[Liao et~al\mbox{.}(2024)]%
        {liao2024llara}
\bibfield{author}{\bibinfo{person}{Jiayi Liao}, \bibinfo{person}{Sihang Li}, \bibinfo{person}{Zhengyi Yang}, \bibinfo{person}{Jiancan Wu}, \bibinfo{person}{Yancheng Yuan}, \bibinfo{person}{Xiang Wang}, {and} \bibinfo{person}{Xiangnan He}.} \bibinfo{year}{2024}\natexlab{}.
\newblock \showarticletitle{Llara: Large language-recommendation assistant}. In \bibinfo{booktitle}{\emph{Proceedings of the 47th International ACM SIGIR Conference on Research and Development in Information Retrieval}}. \bibinfo{pages}{1785--1795}.
\newblock


\bibitem[Lin et~al\mbox{.}(2024a)]%
        {lin2024clickprompt}
\bibfield{author}{\bibinfo{person}{Jianghao Lin}, \bibinfo{person}{Bo Chen}, \bibinfo{person}{Hangyu Wang}, \bibinfo{person}{Yunjia Xi}, \bibinfo{person}{Yanru Qu}, \bibinfo{person}{Xinyi Dai}, \bibinfo{person}{Kangning Zhang}, \bibinfo{person}{Ruiming Tang}, \bibinfo{person}{Yong Yu}, {and} \bibinfo{person}{Weinan Zhang}.} \bibinfo{year}{2024}\natexlab{a}.
\newblock \showarticletitle{ClickPrompt: CTR Models are Strong Prompt Generators for Adapting Language Models to CTR Prediction}. In \bibinfo{booktitle}{\emph{Proceedings of the ACM on Web Conference 2024}}. \bibinfo{pages}{3319--3330}.
\newblock


\bibitem[Lin et~al\mbox{.}(2024b)]%
        {lin2024rella}
\bibfield{author}{\bibinfo{person}{Jianghao Lin}, \bibinfo{person}{Rong Shan}, \bibinfo{person}{Chenxu Zhu}, \bibinfo{person}{Kounianhua Du}, \bibinfo{person}{Bo Chen}, \bibinfo{person}{Shigang Quan}, \bibinfo{person}{Ruiming Tang}, \bibinfo{person}{Yong Yu}, {and} \bibinfo{person}{Weinan Zhang}.} \bibinfo{year}{2024}\natexlab{b}.
\newblock \showarticletitle{Rella: Retrieval-enhanced large language models for lifelong sequential behavior comprehension in recommendation}. In \bibinfo{booktitle}{\emph{Proceedings of the ACM on Web Conference 2024}}. \bibinfo{pages}{3497--3508}.
\newblock


\bibitem[Lin et~al\mbox{.}(2024c)]%
        {lin2024bridging}
\bibfield{author}{\bibinfo{person}{Xinyu Lin}, \bibinfo{person}{Wenjie Wang}, \bibinfo{person}{Yongqi Li}, \bibinfo{person}{Fuli Feng}, \bibinfo{person}{See-Kiong Ng}, {and} \bibinfo{person}{Tat-Seng Chua}.} \bibinfo{year}{2024}\natexlab{c}.
\newblock \showarticletitle{Bridging items and language: A transition paradigm for large language model-based recommendation}. In \bibinfo{booktitle}{\emph{Proceedings of the 30th ACM SIGKDD Conference on Knowledge Discovery and Data Mining}}. \bibinfo{pages}{1816--1826}.
\newblock


\bibitem[Liu et~al\mbox{.}(2023a)]%
        {liu2023recommendation}
\bibfield{author}{\bibinfo{person}{Jiahao Liu}, \bibinfo{person}{Dongsheng Li}, \bibinfo{person}{Hansu Gu}, \bibinfo{person}{Tun Lu}, \bibinfo{person}{Jiongran Wu}, \bibinfo{person}{Peng Zhang}, \bibinfo{person}{Li Shang}, {and} \bibinfo{person}{Ning Gu}.} \bibinfo{year}{2023}\natexlab{a}.
\newblock \showarticletitle{Recommendation unlearning via matrix correction}.
\newblock \bibinfo{journal}{\emph{arXiv preprint arXiv:2307.15960}} (\bibinfo{year}{2023}).
\newblock


\bibitem[Liu et~al\mbox{.}(2022)]%
        {liu2022parameter}
\bibfield{author}{\bibinfo{person}{Jiahao Liu}, \bibinfo{person}{Dongsheng Li}, \bibinfo{person}{Hansu Gu}, \bibinfo{person}{Tun Lu}, \bibinfo{person}{Peng Zhang}, {and} \bibinfo{person}{Ning Gu}.} \bibinfo{year}{2022}\natexlab{}.
\newblock \showarticletitle{Parameter-free dynamic graph embedding for link prediction}.
\newblock \bibinfo{journal}{\emph{Advances in Neural Information Processing Systems}}  \bibinfo{volume}{35} (\bibinfo{year}{2022}), \bibinfo{pages}{27623--27635}.
\newblock


\bibitem[Liu et~al\mbox{.}(2023b)]%
        {liu2023personalized}
\bibfield{author}{\bibinfo{person}{Jiahao Liu}, \bibinfo{person}{Dongsheng Li}, \bibinfo{person}{Hansu Gu}, \bibinfo{person}{Tun Lu}, \bibinfo{person}{Peng Zhang}, \bibinfo{person}{Li Shang}, {and} \bibinfo{person}{Ning Gu}.} \bibinfo{year}{2023}\natexlab{b}.
\newblock \showarticletitle{Personalized graph signal processing for collaborative filtering}. In \bibinfo{booktitle}{\emph{Proceedings of the ACM Web Conference 2023}}. \bibinfo{pages}{1264--1272}.
\newblock


\bibitem[Liu et~al\mbox{.}(2023c)]%
        {liu2023triple}
\bibfield{author}{\bibinfo{person}{Jiahao Liu}, \bibinfo{person}{Dongsheng Li}, \bibinfo{person}{Hansu Gu}, \bibinfo{person}{Tun Lu}, \bibinfo{person}{Peng Zhang}, \bibinfo{person}{Li Shang}, {and} \bibinfo{person}{Ning Gu}.} \bibinfo{year}{2023}\natexlab{c}.
\newblock \showarticletitle{Triple structural information modelling for accurate, explainable and interactive recommendation}. In \bibinfo{booktitle}{\emph{Proceedings of the 46th International ACM SIGIR Conference on Research and Development in Information Retrieval}}. \bibinfo{pages}{1086--1095}.
\newblock


\bibitem[Liu et~al\mbox{.}(2024a)]%
        {liu2024filtering}
\bibfield{author}{\bibinfo{person}{Jiahao Liu}, \bibinfo{person}{Yiyang Shao}, \bibinfo{person}{Peng Zhang}, \bibinfo{person}{Dongsheng Li}, \bibinfo{person}{Hansu Gu}, \bibinfo{person}{Chao Chen}, \bibinfo{person}{Longzhi Du}, \bibinfo{person}{Tun Lu}, {and} \bibinfo{person}{Ning Gu}.} \bibinfo{year}{2024}\natexlab{a}.
\newblock \showarticletitle{Filtering Discomforting Recommendations with Large Language Models}.
\newblock \bibinfo{journal}{\emph{arXiv preprint arXiv:2410.05411}} (\bibinfo{year}{2024}).
\newblock


\bibitem[Liu et~al\mbox{.}(2024b)]%
        {liu2024large}
\bibfield{author}{\bibinfo{person}{Qidong Liu}, \bibinfo{person}{Xian Wu}, \bibinfo{person}{Xiangyu Zhao}, \bibinfo{person}{Yejing Wang}, \bibinfo{person}{Zijian Zhang}, \bibinfo{person}{Feng Tian}, {and} \bibinfo{person}{Yefeng Zheng}.} \bibinfo{year}{2024}\natexlab{b}.
\newblock \showarticletitle{Large Language Models Enhanced Sequential Recommendation for Long-tail User and Item}. In \bibinfo{booktitle}{\emph{Advances in Neural Information Processing Systems}}. \bibinfo{pages}{26701--26727}.
\newblock


\bibitem[Liu et~al\mbox{.}(2024c)]%
        {liu2024largesur}
\bibfield{author}{\bibinfo{person}{Qidong Liu}, \bibinfo{person}{Xiangyu Zhao}, \bibinfo{person}{Yuhao Wang}, \bibinfo{person}{Yejing Wang}, \bibinfo{person}{Zijian Zhang}, \bibinfo{person}{Yuqi Sun}, \bibinfo{person}{Xiang Li}, \bibinfo{person}{Maolin Wang}, \bibinfo{person}{Pengyue Jia}, \bibinfo{person}{Chong Chen}, {et~al\mbox{.}}} \bibinfo{year}{2024}\natexlab{c}.
\newblock \showarticletitle{Large Language Model Enhanced Recommender Systems: Taxonomy, Trend, Application and Future}.
\newblock \bibinfo{journal}{\emph{arXiv preprint arXiv:2412.13432}} (\bibinfo{year}{2024}).
\newblock


\bibitem[Liu et~al\mbox{.}(2023d)]%
        {liu2023autoseqrec}
\bibfield{author}{\bibinfo{person}{Sijia Liu}, \bibinfo{person}{Jiahao Liu}, \bibinfo{person}{Hansu Gu}, \bibinfo{person}{Dongsheng Li}, \bibinfo{person}{Tun Lu}, \bibinfo{person}{Peng Zhang}, {and} \bibinfo{person}{Ning Gu}.} \bibinfo{year}{2023}\natexlab{d}.
\newblock \showarticletitle{Autoseqrec: Autoencoder for efficient sequential recommendation}. In \bibinfo{booktitle}{\emph{Proceedings of the 32nd ACM International Conference on Information and Knowledge Management}}. \bibinfo{pages}{1493--1502}.
\newblock


\bibitem[Lu et~al\mbox{.}(2024)]%
        {lu2024aligning}
\bibfield{author}{\bibinfo{person}{Wensheng Lu}, \bibinfo{person}{Jianxun Lian}, \bibinfo{person}{Wei Zhang}, \bibinfo{person}{Guanghua Li}, \bibinfo{person}{Mingyang Zhou}, \bibinfo{person}{Hao Liao}, {and} \bibinfo{person}{Xing Xie}.} \bibinfo{year}{2024}\natexlab{}.
\newblock \showarticletitle{Aligning Large Language Models for Controllable Recommendations}. In \bibinfo{booktitle}{\emph{Proceedings of the 62nd Annual Meeting of the Association for Computational Linguistics (Volume 1: Long Papers)}}. \bibinfo{pages}{8159--8172}.
\newblock


\bibitem[Luo et~al\mbox{.}(2024b)]%
        {luo2024recranker}
\bibfield{author}{\bibinfo{person}{Sichun Luo}, \bibinfo{person}{Bowei He}, \bibinfo{person}{Haohan Zhao}, \bibinfo{person}{Wei Shao}, \bibinfo{person}{Yanlin Qi}, \bibinfo{person}{Yinya Huang}, \bibinfo{person}{Aojun Zhou}, \bibinfo{person}{Yuxuan Yao}, \bibinfo{person}{Zongpeng Li}, \bibinfo{person}{Yuanzhang Xiao}, {et~al\mbox{.}}} \bibinfo{year}{2024}\natexlab{b}.
\newblock \showarticletitle{Recranker: Instruction tuning large language model as ranker for top-k recommendation}.
\newblock \bibinfo{journal}{\emph{ACM Transactions on Information Systems}} (\bibinfo{year}{2024}).
\newblock


\bibitem[Luo et~al\mbox{.}(2024a)]%
        {luo2024unlocking}
\bibfield{author}{\bibinfo{person}{Yucong Luo}, \bibinfo{person}{Mingyue Cheng}, \bibinfo{person}{Hao Zhang}, \bibinfo{person}{Junyu Lu}, {and} \bibinfo{person}{Enhong Chen}.} \bibinfo{year}{2024}\natexlab{a}.
\newblock \showarticletitle{Unlocking the potential of large language models for explainable recommendations}. In \bibinfo{booktitle}{\emph{International Conference on Database Systems for Advanced Applications}}. Springer, \bibinfo{pages}{286--303}.
\newblock


\bibitem[Medsker et~al\mbox{.}(2001)]%
        {medsker2001recurrent}
\bibfield{author}{\bibinfo{person}{Larry~R Medsker}, \bibinfo{person}{Lakhmi Jain}, {et~al\mbox{.}}} \bibinfo{year}{2001}\natexlab{}.
\newblock \showarticletitle{Recurrent neural networks}.
\newblock \bibinfo{journal}{\emph{Design and Applications}} \bibinfo{volume}{5}, \bibinfo{number}{64-67} (\bibinfo{year}{2001}), \bibinfo{pages}{2}.
\newblock


\bibitem[Petruzzelli et~al\mbox{.}(2024)]%
        {petruzzelli2024instructing}
\bibfield{author}{\bibinfo{person}{Alessandro Petruzzelli}, \bibinfo{person}{Cataldo Musto}, \bibinfo{person}{Lucrezia Laraspata}, \bibinfo{person}{Ivan Rinaldi}, \bibinfo{person}{Marco de Gemmis}, \bibinfo{person}{Pasquale Lops}, {and} \bibinfo{person}{Giovanni Semeraro}.} \bibinfo{year}{2024}\natexlab{}.
\newblock \showarticletitle{Instructing and prompting large language models for explainable cross-domain recommendations}. In \bibinfo{booktitle}{\emph{Proceedings of the 18th ACM Conference on Recommender Systems}}. \bibinfo{pages}{298--308}.
\newblock


\bibitem[Ren et~al\mbox{.}(2024)]%
        {ren2024representation}
\bibfield{author}{\bibinfo{person}{Xubin Ren}, \bibinfo{person}{Wei Wei}, \bibinfo{person}{Lianghao Xia}, \bibinfo{person}{Lixin Su}, \bibinfo{person}{Suqi Cheng}, \bibinfo{person}{Junfeng Wang}, \bibinfo{person}{Dawei Yin}, {and} \bibinfo{person}{Chao Huang}.} \bibinfo{year}{2024}\natexlab{}.
\newblock \showarticletitle{Representation learning with large language models for recommendation}. In \bibinfo{booktitle}{\emph{Proceedings of the ACM on Web Conference 2024}}. \bibinfo{pages}{3464--3475}.
\newblock


\bibitem[Sakurai et~al\mbox{.}(2024)]%
        {sakurai2024llm}
\bibfield{author}{\bibinfo{person}{Keigo Sakurai}, \bibinfo{person}{Ren Togo}, \bibinfo{person}{Takahiro Ogawa}, {and} \bibinfo{person}{Miki Haseyama}.} \bibinfo{year}{2024}\natexlab{}.
\newblock \showarticletitle{LLM is Knowledge Graph Reasoner: LLM's Intuition-aware Knowledge Graph Reasoning for Cold-start Sequential Recommendation}.
\newblock \bibinfo{journal}{\emph{arXiv preprint arXiv:2412.12464}} (\bibinfo{year}{2024}).
\newblock


\bibitem[Sun et~al\mbox{.}(2024a)]%
        {sun2024adaptive}
\bibfield{author}{\bibinfo{person}{Zhu Sun}, \bibinfo{person}{Kaidong Feng}, \bibinfo{person}{Jie Yang}, \bibinfo{person}{Xinghua Qu}, \bibinfo{person}{Hui Fang}, \bibinfo{person}{Yew-Soon Ong}, {and} \bibinfo{person}{Wenyuan Liu}.} \bibinfo{year}{2024}\natexlab{a}.
\newblock \showarticletitle{Adaptive In-Context Learning with Large Language Models for Bundle Generation}. In \bibinfo{booktitle}{\emph{Proceedings of the 47th International ACM SIGIR Conference on Research and Development in Information Retrieval}}. \bibinfo{pages}{966--976}.
\newblock


\bibitem[Sun et~al\mbox{.}(2024b)]%
        {sun2024large}
\bibfield{author}{\bibinfo{person}{Zhu Sun}, \bibinfo{person}{Hongyang Liu}, \bibinfo{person}{Xinghua Qu}, \bibinfo{person}{Kaidong Feng}, \bibinfo{person}{Yan Wang}, {and} \bibinfo{person}{Yew~Soon Ong}.} \bibinfo{year}{2024}\natexlab{b}.
\newblock \showarticletitle{Large language models for intent-driven session recommendations}. In \bibinfo{booktitle}{\emph{Proceedings of the 47th International ACM SIGIR Conference on Research and Development in Information Retrieval}}. \bibinfo{pages}{324--334}.
\newblock


\bibitem[Sun et~al\mbox{.}(2024c)]%
        {sun2024largecf}
\bibfield{author}{\bibinfo{person}{Zhongxiang Sun}, \bibinfo{person}{Zihua Si}, \bibinfo{person}{Xiaoxue Zang}, \bibinfo{person}{Kai Zheng}, \bibinfo{person}{Yang Song}, \bibinfo{person}{Xiao Zhang}, {and} \bibinfo{person}{Jun Xu}.} \bibinfo{year}{2024}\natexlab{c}.
\newblock \showarticletitle{Large language models enhanced collaborative filtering}. In \bibinfo{booktitle}{\emph{Proceedings of the 33rd ACM International Conference on Information and Knowledge Management}}. \bibinfo{pages}{2178--2188}.
\newblock


\bibitem[Wang et~al\mbox{.}(2024b)]%
        {wang2024large}
\bibfield{author}{\bibinfo{person}{Jianling Wang}, \bibinfo{person}{Haokai Lu}, \bibinfo{person}{James Caverlee}, \bibinfo{person}{Ed~H Chi}, {and} \bibinfo{person}{Minmin Chen}.} \bibinfo{year}{2024}\natexlab{b}.
\newblock \showarticletitle{Large Language Models as Data Augmenters for Cold-Start Item Recommendation}. In \bibinfo{booktitle}{\emph{Companion Proceedings of the ACM on Web Conference 2024}}. \bibinfo{pages}{726--729}.
\newblock


\bibitem[Wang et~al\mbox{.}(2024c)]%
        {wang2024llms}
\bibfield{author}{\bibinfo{person}{Jianling Wang}, \bibinfo{person}{Haokai Lu}, \bibinfo{person}{Yifan Liu}, \bibinfo{person}{He Ma}, \bibinfo{person}{Yueqi Wang}, \bibinfo{person}{Yang Gu}, \bibinfo{person}{Shuzhou Zhang}, \bibinfo{person}{Ningren Han}, \bibinfo{person}{Shuchao Bi}, \bibinfo{person}{Lexi Baugher}, {et~al\mbox{.}}} \bibinfo{year}{2024}\natexlab{c}.
\newblock \showarticletitle{Llms for user interest exploration in large-scale recommendation systems}. In \bibinfo{booktitle}{\emph{Proceedings of the 18th ACM Conference on Recommender Systems}}. \bibinfo{pages}{872--877}.
\newblock


\bibitem[Wang and Lim(2024)]%
        {wang2024whole}
\bibfield{author}{\bibinfo{person}{Lei Wang} {and} \bibinfo{person}{Ee-Peng Lim}.} \bibinfo{year}{2024}\natexlab{}.
\newblock \showarticletitle{The Whole is Better than the Sum: Using Aggregated Demonstrations in In-Context Learning for Sequential Recommendation}. In \bibinfo{booktitle}{\emph{Findings of the Association for Computational Linguistics: NAACL 2024}}. \bibinfo{pages}{876--895}.
\newblock


\bibitem[Wang et~al\mbox{.}(2024a)]%
        {wang2024llmrg}
\bibfield{author}{\bibinfo{person}{Yan Wang}, \bibinfo{person}{Zhixuan Chu}, \bibinfo{person}{Xin Ouyang}, \bibinfo{person}{Simeng Wang}, \bibinfo{person}{Hongyan Hao}, \bibinfo{person}{Yue Shen}, \bibinfo{person}{Jinjie Gu}, \bibinfo{person}{Siqiao Xue}, \bibinfo{person}{James Zhang}, \bibinfo{person}{Qing Cui}, {et~al\mbox{.}}} \bibinfo{year}{2024}\natexlab{a}.
\newblock \showarticletitle{LLMRG: Improving Recommendations through Large Language Model Reasoning Graphs}. In \bibinfo{booktitle}{\emph{Proceedings of the AAAI Conference on Artificial Intelligence}}, Vol.~\bibinfo{volume}{38}. \bibinfo{pages}{19189--19196}.
\newblock


\bibitem[Wang et~al\mbox{.}(2024d)]%
        {wang2024can}
\bibfield{author}{\bibinfo{person}{Yuling Wang}, \bibinfo{person}{Changxin Tian}, \bibinfo{person}{Binbin Hu}, \bibinfo{person}{Yanhua Yu}, \bibinfo{person}{Ziqi Liu}, \bibinfo{person}{Zhiqiang Zhang}, \bibinfo{person}{Jun Zhou}, \bibinfo{person}{Liang Pang}, {and} \bibinfo{person}{Xiao Wang}.} \bibinfo{year}{2024}\natexlab{d}.
\newblock \showarticletitle{Can Small Language Models be Good Reasoners for Sequential Recommendation?}. In \bibinfo{booktitle}{\emph{Proceedings of the ACM on Web Conference 2024}}. \bibinfo{pages}{3876--3887}.
\newblock


\bibitem[Wei et~al\mbox{.}(2022)]%
        {wei2022chain}
\bibfield{author}{\bibinfo{person}{Jason Wei}, \bibinfo{person}{Xuezhi Wang}, \bibinfo{person}{Dale Schuurmans}, \bibinfo{person}{Maarten Bosma}, \bibinfo{person}{Fei Xia}, \bibinfo{person}{Ed Chi}, \bibinfo{person}{Quoc~V Le}, \bibinfo{person}{Denny Zhou}, {et~al\mbox{.}}} \bibinfo{year}{2022}\natexlab{}.
\newblock \showarticletitle{Chain-of-thought prompting elicits reasoning in large language models}.
\newblock \bibinfo{journal}{\emph{Advances in neural information processing systems}}  \bibinfo{volume}{35} (\bibinfo{year}{2022}), \bibinfo{pages}{24824--24837}.
\newblock


\bibitem[Wei et~al\mbox{.}(2024)]%
        {wei2024llmrec}
\bibfield{author}{\bibinfo{person}{Wei Wei}, \bibinfo{person}{Xubin Ren}, \bibinfo{person}{Jiabin Tang}, \bibinfo{person}{Qinyong Wang}, \bibinfo{person}{Lixin Su}, \bibinfo{person}{Suqi Cheng}, \bibinfo{person}{Junfeng Wang}, \bibinfo{person}{Dawei Yin}, {and} \bibinfo{person}{Chao Huang}.} \bibinfo{year}{2024}\natexlab{}.
\newblock \showarticletitle{Llmrec: Large language models with graph augmentation for recommendation}. In \bibinfo{booktitle}{\emph{Proceedings of the 17th ACM International Conference on Web Search and Data Mining}}. \bibinfo{pages}{806--815}.
\newblock


\bibitem[Wu et~al\mbox{.}(2020)]%
        {wu2020mind}
\bibfield{author}{\bibinfo{person}{Fangzhao Wu}, \bibinfo{person}{Ying Qiao}, \bibinfo{person}{Jiun-Hung Chen}, \bibinfo{person}{Chuhan Wu}, \bibinfo{person}{Tao Qi}, \bibinfo{person}{Jianxun Lian}, \bibinfo{person}{Danyang Liu}, \bibinfo{person}{Xing Xie}, \bibinfo{person}{Jianfeng Gao}, \bibinfo{person}{Winnie Wu}, {et~al\mbox{.}}} \bibinfo{year}{2020}\natexlab{}.
\newblock \showarticletitle{Mind: A large-scale dataset for news recommendation}. In \bibinfo{booktitle}{\emph{Proceedings of the 58th annual meeting of the association for computational linguistics}}. \bibinfo{pages}{3597--3606}.
\newblock


\bibitem[Xi et~al\mbox{.}(2024)]%
        {xi2024towards}
\bibfield{author}{\bibinfo{person}{Yunjia Xi}, \bibinfo{person}{Weiwen Liu}, \bibinfo{person}{Jianghao Lin}, \bibinfo{person}{Xiaoling Cai}, \bibinfo{person}{Hong Zhu}, \bibinfo{person}{Jieming Zhu}, \bibinfo{person}{Bo Chen}, \bibinfo{person}{Ruiming Tang}, \bibinfo{person}{Weinan Zhang}, {and} \bibinfo{person}{Yong Yu}.} \bibinfo{year}{2024}\natexlab{}.
\newblock \showarticletitle{Towards open-world recommendation with knowledge augmentation from large language models}. In \bibinfo{booktitle}{\emph{Proceedings of the 18th ACM Conference on Recommender Systems}}. \bibinfo{pages}{12--22}.
\newblock


\bibitem[Xia et~al\mbox{.}(2022)]%
        {xia2022fire}
\bibfield{author}{\bibinfo{person}{Jiafeng Xia}, \bibinfo{person}{Dongsheng Li}, \bibinfo{person}{Hansu Gu}, \bibinfo{person}{Jiahao Liu}, \bibinfo{person}{Tun Lu}, {and} \bibinfo{person}{Ning Gu}.} \bibinfo{year}{2022}\natexlab{}.
\newblock \showarticletitle{FIRE: Fast incremental recommendation with graph signal processing}. In \bibinfo{booktitle}{\emph{Proceedings of the ACM Web Conference 2022}}. \bibinfo{pages}{2360--2369}.
\newblock


\bibitem[Yang et~al\mbox{.}(2024)]%
        {yang2024sequential}
\bibfield{author}{\bibinfo{person}{Shenghao Yang}, \bibinfo{person}{Weizhi Ma}, \bibinfo{person}{Peijie Sun}, \bibinfo{person}{Qingyao Ai}, \bibinfo{person}{Yiqun Liu}, \bibinfo{person}{Mingchen Cai}, {and} \bibinfo{person}{Min Zhang}.} \bibinfo{year}{2024}\natexlab{}.
\newblock \showarticletitle{Sequential recommendation with latent relations based on large language model}. In \bibinfo{booktitle}{\emph{Proceedings of the 47th International ACM SIGIR Conference on Research and Development in Information Retrieval}}. \bibinfo{pages}{335--344}.
\newblock


\bibitem[Yu et~al\mbox{.}(2024)]%
        {yu2024break}
\bibfield{author}{\bibinfo{person}{Xiaohan Yu}, \bibinfo{person}{Li Zhang}, \bibinfo{person}{Xin Zhao}, {and} \bibinfo{person}{Yue Wang}.} \bibinfo{year}{2024}\natexlab{}.
\newblock \showarticletitle{Break the ID-Language Barrier: An Adaption Framework for Sequential Recommendation}.
\newblock \bibinfo{journal}{\emph{arXiv preprint arXiv:2411.18262}} (\bibinfo{year}{2024}).
\newblock


\bibitem[Yue et~al\mbox{.}(2023)]%
        {yue2023llamarec}
\bibfield{author}{\bibinfo{person}{Zhenrui Yue}, \bibinfo{person}{Sara Rabhi}, \bibinfo{person}{Gabriel de Souza~Pereira Moreira}, \bibinfo{person}{Dong Wang}, {and} \bibinfo{person}{Even Oldridge}.} \bibinfo{year}{2023}\natexlab{}.
\newblock \showarticletitle{LlamaRec: Two-stage recommendation using large language models for ranking}.
\newblock \bibinfo{journal}{\emph{arXiv preprint arXiv:2311.02089}} (\bibinfo{year}{2023}).
\newblock


\bibitem[Zhang et~al\mbox{.}(2024d)]%
        {zhang2024embsum}
\bibfield{author}{\bibinfo{person}{Chiyu Zhang}, \bibinfo{person}{Yifei Sun}, \bibinfo{person}{Minghao Wu}, \bibinfo{person}{Jun Chen}, \bibinfo{person}{Jie Lei}, \bibinfo{person}{Muhammad Abdul-Mageed}, \bibinfo{person}{Rong Jin}, \bibinfo{person}{Angli Liu}, \bibinfo{person}{Ji Zhu}, \bibinfo{person}{Sem Park}, {et~al\mbox{.}}} \bibinfo{year}{2024}\natexlab{d}.
\newblock \showarticletitle{Embsum: Leveraging the summarization capabilities of large language models for content-based recommendations}. In \bibinfo{booktitle}{\emph{Proceedings of the 18th ACM Conference on Recommender Systems}}. \bibinfo{pages}{1010--1015}.
\newblock


\bibitem[Zhang et~al\mbox{.}(2024e)]%
        {zhang2024notellm}
\bibfield{author}{\bibinfo{person}{Chao Zhang}, \bibinfo{person}{Shiwei Wu}, \bibinfo{person}{Haoxin Zhang}, \bibinfo{person}{Tong Xu}, \bibinfo{person}{Yan Gao}, \bibinfo{person}{Yao Hu}, {and} \bibinfo{person}{Enhong Chen}.} \bibinfo{year}{2024}\natexlab{e}.
\newblock \showarticletitle{NoteLLM: A Retrievable Large Language Model for Note Recommendation}. In \bibinfo{booktitle}{\emph{Companion Proceedings of the ACM on Web Conference 2024}}. \bibinfo{pages}{170--179}.
\newblock


\bibitem[Zhang et~al\mbox{.}(2023)]%
        {zhang2023chatgpt}
\bibfield{author}{\bibinfo{person}{Jizhi Zhang}, \bibinfo{person}{Keqin Bao}, \bibinfo{person}{Yang Zhang}, \bibinfo{person}{Wenjie Wang}, \bibinfo{person}{Fuli Feng}, {and} \bibinfo{person}{Xiangnan He}.} \bibinfo{year}{2023}\natexlab{}.
\newblock \showarticletitle{Is chatgpt fair for recommendation? evaluating fairness in large language model recommendation}. In \bibinfo{booktitle}{\emph{Proceedings of the 17th ACM Conference on Recommender Systems}}. \bibinfo{pages}{993--999}.
\newblock


\bibitem[Zhang et~al\mbox{.}(2024b)]%
        {zhang2024agentcf}
\bibfield{author}{\bibinfo{person}{Junjie Zhang}, \bibinfo{person}{Yupeng Hou}, \bibinfo{person}{Ruobing Xie}, \bibinfo{person}{Wenqi Sun}, \bibinfo{person}{Julian McAuley}, \bibinfo{person}{Wayne~Xin Zhao}, \bibinfo{person}{Leyu Lin}, {and} \bibinfo{person}{Ji-Rong Wen}.} \bibinfo{year}{2024}\natexlab{b}.
\newblock \showarticletitle{Agentcf: Collaborative learning with autonomous language agents for recommender systems}. In \bibinfo{booktitle}{\emph{Proceedings of the ACM on Web Conference 2024}}. \bibinfo{pages}{3679--3689}.
\newblock


\bibitem[Zhang et~al\mbox{.}(2024c)]%
        {zhang2024large}
\bibfield{author}{\bibinfo{person}{Xiaoyu Zhang}, \bibinfo{person}{Yishan Li}, \bibinfo{person}{Jiayin Wang}, \bibinfo{person}{Bowen Sun}, \bibinfo{person}{Weizhi Ma}, \bibinfo{person}{Peijie Sun}, {and} \bibinfo{person}{Min Zhang}.} \bibinfo{year}{2024}\natexlab{c}.
\newblock \showarticletitle{Large language models as evaluators for recommendation explanations}. In \bibinfo{booktitle}{\emph{Proceedings of the 18th ACM Conference on Recommender Systems}}. \bibinfo{pages}{33--42}.
\newblock


\bibitem[Zhang et~al\mbox{.}(2024f)]%
        {zhang2024finerec}
\bibfield{author}{\bibinfo{person}{Xiaokun Zhang}, \bibinfo{person}{Bo Xu}, \bibinfo{person}{Youlin Wu}, \bibinfo{person}{Yuan Zhong}, \bibinfo{person}{Hongfei Lin}, {and} \bibinfo{person}{Fenglong Ma}.} \bibinfo{year}{2024}\natexlab{f}.
\newblock \showarticletitle{Finerec: Exploring fine-grained sequential recommendation}. In \bibinfo{booktitle}{\emph{Proceedings of the 47th International ACM SIGIR Conference on Research and Development in Information Retrieval}}. \bibinfo{pages}{1599--1608}.
\newblock


\bibitem[Zhang et~al\mbox{.}(2024a)]%
        {zhang2024text}
\bibfield{author}{\bibinfo{person}{Yang Zhang}, \bibinfo{person}{Keqin Bao}, \bibinfo{person}{Ming Yan}, \bibinfo{person}{Wenjie Wang}, \bibinfo{person}{Fuli Feng}, {and} \bibinfo{person}{Xiangnan He}.} \bibinfo{year}{2024}\natexlab{a}.
\newblock \showarticletitle{Text-like Encoding of Collaborative Information in Large Language Models for Recommendation}. In \bibinfo{booktitle}{\emph{Proceedings of the 62nd Annual Meeting of the Association for Computational Linguistics (Volume 1: Long Papers)}}. \bibinfo{pages}{9181--9191}.
\newblock


\bibitem[Zhang et~al\mbox{.}(2024g)]%
        {zhang2024recgpt}
\bibfield{author}{\bibinfo{person}{Yabin Zhang}, \bibinfo{person}{Wenhui Yu}, \bibinfo{person}{Erhan Zhang}, \bibinfo{person}{Xu Chen}, \bibinfo{person}{Lantao Hu}, \bibinfo{person}{Peng Jiang}, {and} \bibinfo{person}{Kun Gai}.} \bibinfo{year}{2024}\natexlab{g}.
\newblock \showarticletitle{RecGPT: Generative Personalized Prompts for Sequential Recommendation via ChatGPT Training Paradigm}.
\newblock \bibinfo{journal}{\emph{arXiv preprint arXiv:2404.08675}} (\bibinfo{year}{2024}).
\newblock


\bibitem[Zhao et~al\mbox{.}(2024)]%
        {zhao2024let}
\bibfield{author}{\bibinfo{person}{Yuyue Zhao}, \bibinfo{person}{Jiancan Wu}, \bibinfo{person}{Xiang Wang}, \bibinfo{person}{Wei Tang}, \bibinfo{person}{Dingxian Wang}, {and} \bibinfo{person}{Maarten De~Rijke}.} \bibinfo{year}{2024}\natexlab{}.
\newblock \showarticletitle{Let me do it for you: Towards llm empowered recommendation via tool learning}. In \bibinfo{booktitle}{\emph{Proceedings of the 47th International ACM SIGIR Conference on Research and Development in Information Retrieval}}. \bibinfo{pages}{1796--1806}.
\newblock


\bibitem[Zheng et~al\mbox{.}(2024b)]%
        {zheng2024adapting}
\bibfield{author}{\bibinfo{person}{Bowen Zheng}, \bibinfo{person}{Yupeng Hou}, \bibinfo{person}{Hongyu Lu}, \bibinfo{person}{Yu Chen}, \bibinfo{person}{Wayne~Xin Zhao}, \bibinfo{person}{Ming Chen}, {and} \bibinfo{person}{Ji-Rong Wen}.} \bibinfo{year}{2024}\natexlab{b}.
\newblock \showarticletitle{Adapting large language models by integrating collaborative semantics for recommendation}. In \bibinfo{booktitle}{\emph{2024 IEEE 40th International Conference on Data Engineering (ICDE)}}. IEEE, \bibinfo{pages}{1435--1448}.
\newblock


\bibitem[Zheng et~al\mbox{.}(2024a)]%
        {zheng2024harnessing}
\bibfield{author}{\bibinfo{person}{Zhi Zheng}, \bibinfo{person}{Wenshuo Chao}, \bibinfo{person}{Zhaopeng Qiu}, \bibinfo{person}{Hengshu Zhu}, {and} \bibinfo{person}{Hui Xiong}.} \bibinfo{year}{2024}\natexlab{a}.
\newblock \showarticletitle{Harnessing large language models for text-rich sequential recommendation}. In \bibinfo{booktitle}{\emph{Proceedings of the ACM on Web Conference 2024}}. \bibinfo{pages}{3207--3216}.
\newblock


\end{thebibliography}

\appendix

\end{document}